# Light Bullets in Su–Schrieffer–Heeger Photonic Topological Insulators

Sergey K. Ivanov,[1] Yaroslav V. Kartashov,[1,2] and Lluis Torner[1,3]

[1]*ICFO-Institut de Ciencies Fotoniques, The Barcelona Institute of Science and Technology, 08860 Castelldefels (Barcelona), Spain*
[2]*Institute of Spectroscopy, Russian Academy of Sciences, 108840, Troitsk, Moscow, Russia*
[3]*Universitat Politecnica de Catalunya 08034 Barcelona, Spain*
*Corresponding author: sergey.ivanov@icfo.eu

We introduce a different class of thresholdless three-dimensional soliton states that form in higher-order topological insulators based on a two-dimensional Su–Schrieffer–Heeger array of coupled waveguides. The linear spectrum of such structures is characterized by the presence of a topological gap with corner states residing in them. We find that a focusing Kerr nonlinearity allows families of light bullets bifurcating from the linear corner states to exist as stable three-dimensional solitons, which inherit topological protection from their linear corner counterparts and, remarkably, survive even in the presence of considerable disorder. The light bullets exhibit a spatial localization degree that depends strongly on the array dimerization, and may feature large temporal widths in the topological gap near the bifurcation point, thus drastically reducing the otherwise strong instabilities caused by higher-order effects.

After the original discovery in condensed matter, topological insulators have been encountered in several other areas of physics, leading to their experimental demonstration in many systems [1-3], including optical settings [4-6]. An important property of topological insulators is the existence of topologically protected edge states with energies residing inside a topological gap. Recently, a different class of higher-order topological insulators has been suggested [7-10], the most remarkable feature of which is their ability to support topological states with lower dimensions than the bulk [11,12]. Higher-order topological insulators underlie many far-reaching concepts, such as higher-order band topology in twisted moiré superlattices [13], Majorana-like bound states [14] and their nontrivial braiding [15], or topological lattice disclinations [16], to name a few.

Optical systems afford the possibility to combine topological effects and nonlinear self-action, hence enabling a plethora of phenomena, such as modulational instabilities of the topological states [17-19], inversion of topological currents [20,21], nonlinear tuning of the edge state energies [22], induction of topologically nontrivial phases [23-26], enhancement of parametric interactions [27,28], and rich bistability effects [29,30]. Advances in the field are reviewed in [31-33]. In particular, nonlinearity allows the formation of so-called edge solitons — unique states that exhibit topological protection, that appear in a variety of shapes and feature unusual interactions. Thus, Floquet edge solitons in helical waveguide arrays have been theoretically predicted in continuous [18,34-37], discrete [38-41], and Dirac [42] models. Moreover, Bragg [43], multicomponent [36,37], valley-Hall edge solitons [44,45] and solitons in medium with other than Kerr-type nonlinearity [46] have been addressed. Topological optical solitons have recently been observed experimentally [47-50], as well as nonlinear states bifurcating from corner modes in higher-order topological insulators [51,52]. All such states are either one- (1D) or two-dimensional (2D).

Three-dimensional (3D) wave packets, referred to as light bullets [53], are nondiffracting and nondispersing states. They have attracted continuous attention during several decades (see [54-56]). However, in contrast to spatial 1D and 2D solitons, their experimental realization as stable, long-lasting states remains an outstanding open challenge because of practical difficulties to fabricate a suitable material structure that supports stable bullets and also, more fundamentally, because multidimensional solitons are prone to strong instabilities [57]. Various theoretical schemes to realize stable multidimensional states have been proposed over the years [55,56], including parametric mixing in quadratic nonlinear media [58,59], nonlocal [60,61], competing [62,63], and saturable [64] nonlinearities, as well as dissipative effects [65-70]. Transversally modulated, nonlinear media, e.g., arrays of evanescently coupled waveguides, have also been predicted to support stable light bullets [71-74]. The first experimental observation of light bullets in a hexagonal fiber-like array with silica cores was reported in [75], a work that led to the observation of transient fundamental [76,77] and weakly unstable vortex [78] light bullets. Later, nonlinearity-induced locking of long pulses in different modes, resulting in the formation of spatiotemporal localized states, was observed in graded-index multimode optical fibers [79]. The recent progress in the realization of nonlinear photonic topological insulators in periodic systems raises the question of whether they can support stable light bullets of topological origin.

In this paper, we explore a 2D Su–Schrieffer–Heeger (SSH) [80] optical lattice, possessing second-order localized corner modes, to realize families of stable topological corner 3D light bullets bifurcating from linear corner states belonging to a topological gap. Such bullets are localized both in space and time, and their spatial structure can be controlled by changing the dimerization of the array, while their temporal localization depends on their detuning from linear corner states, which allows to obtain well-localized states with temporal durations at which higher-order effects can be neglected. They inherit the staggered spatial structure from corner modes, which distinguishes them from usual light bullets in periodic optical potentials. The bullets do not require energy threshold for their existence, and can be stable at low and high energies, even in weak optical potentials. They are robust against considerable disorder introduced into the underlying waveguide array due to topological protection.

We address the propagation of 3D light beams along the $z$-axis in a medium with a Kerr $\chi^{(3)}$ nonlinearity and an inhomogeneous refractive index distribution forming a 2D SSH array of evanescently coupled waveguides. The corresponding normalized evolution equation for the light field reads as

$$i\frac{\partial \psi}{\partial z} = -\frac{1}{2}\left(\frac{\partial^2 \psi}{\partial x^2} + \frac{\partial^2 \psi}{\partial y^2}\right) - \frac{1}{2}\frac{\partial^2 \psi}{\partial t^2} - |\psi|^2\psi - R(\boldsymbol{r})\psi. \quad (1)$$

Here the transverse coordinates $r = (x, y)$ and the propagation distance $z = Z/Z_d$ are normalized to the characteristic transverse scale $w_0$ and diffraction length $Z_d = k_0 n_0 w_0^2$, respectively, where $k_0 = \omega_0/c$ is the wavenumber, $\omega_0$ is the carrier frequency, $n_0$ is the unperturbed refractive index defining $\kappa(\omega) = n_0(\omega)\omega/c$, $t = (T - Z/v_g)/T_s$ is the time in the frame moving with group velocity $v_g$, $T_s = w_0[-\kappa^{(2)}\kappa(\omega)]^{1/2}$ is the time scaling, $\kappa^{(2)} = \partial_\omega^2 \kappa(\omega) < 0$ is the anomalous group velocity dispersion coefficient. Our 2D SSH array possesses a square geometry with four waveguides in a unit cell as depicted in Fig. 1. There are two competing parameters in the array: the intracell ($d_1$) and the intercell ($d_2$) distances between waveguides. They modulate the coupling strengths between nearest lattice sites, while $d_1 + d_2 = 2d$, where $2d$ is the lattice constant. For simplicity, we define dimerization parameter $\Delta = (d_1 - d_2)/2\sqrt{2}$, that is, the diagonal shift of the waveguides from the equilibrium position when the distance between neighboring waveguides is equal to $d$ (for $\Delta = 0$, the array becomes square with the lattice constant $d$). The array is composed of identical waveguides of width $\sigma$ placed in the nodes $r_{nm} = (x_{nm}, y_{nm})$ of the 2D SSH grid $R(r) = p\sum_{nm} \exp[-(r - r_{nm})^2/\sigma^2]$ with depth $p = \max(\delta n) k_0^2 w_0^2 n_0$, where $\delta n$ is the refractive index contrast. Having in mind a potential realization of our system with fs-laser written waveguide arrays in fused silica [81], we select the values of the dimensionless parameters to be $d = 3, \sigma = 0.4$, and $p = 4$. This corresponds to a 60 $\mu$m distance between waveguides at $\Delta = 0$, a waveguide width of 8 $\mu$m (hereafter, we use the characteristic transverse scale $w_0 = 20$ $\mu$m), a refractive index contrast of about $\delta n \approx 4.2 \times 10^{-4}$ at the wavelength $\lambda = 1550$ nm (the background refractive index is $n_0 \approx 1.45$); while $z = 1$ corresponds to about 2.34 mm. The group velocity dispersion for such a wavelength is of the order of $\kappa^{(2)} \approx -28$ fs$^2$/mm and the time scaling factor is $T_s = 8.1$ fs.

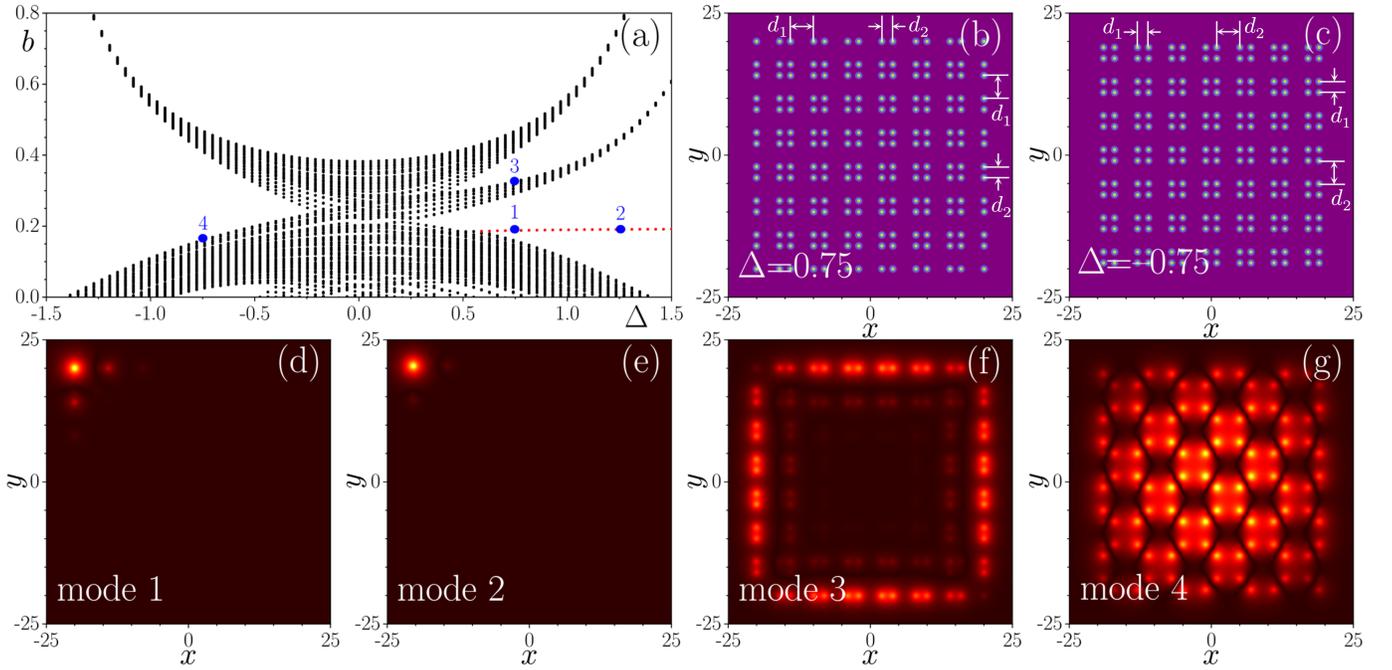

Fig. 1. Propagation constants of the linear modes of the 2D SSH array vs dimerization parameters $\Delta$ (a). The red line corresponds to corner topological modes, the blue dots correspond to the modes shown in the bottom row of the figure. Lattice profiles for $\Delta = 0.75$ (b) and $\Delta = -0.75$ (c). Corner modes for $\Delta = 0.75$ (d) and $\Delta = 1.3$ (e). Edge states for $\Delta = 0.75$ (f). An example of a bulk mode in a nontopological lattice (g) for $\Delta = -0.75$. Here and below $p = 4$, $d = 3$ and $\sigma = 0.4$. Here and in all figures below all quantities are plotted in dimensionless units.

For the 2D SSH array a topological phase can be introduced by varying the dimerization parameter $\Delta$ (see Fig. 1). The examples of arrays with positive and negative $\Delta$ are illustrated in Figs. 1(b) and 1(c). The emergence of a topological phase in the array of this type can be characterized by the corresponding topological invariants defined for the periodic non-truncated array — two polarizations

$$P_j = -S^{-1} \iint_{BZ} \sum_l A_j^{ll} \, dk_x dk_y, \qquad (2)$$

where $A_j^{nm} = -i\langle \phi_{k,n} | \partial \phi_{k,m}/\partial k_j \rangle$ is the Berry connection with product $\langle u | g \rangle = \iint_{unit\ cell} u^* g \, dxdy$, $S$ is the area of the first Brillouin zone (BZ), $\mathbf{k} = (k_x, k_y)$ is Bloch momentum, $\psi_n = \phi_{k,n} \exp(i\mathbf{k}\mathbf{r} + ibz)$, where $\phi_{k,n}(x, y) = \phi_{k,n}(x + 2d, y + 2d)$ is the Bloch function of $n$th band which is periodic along $x$ and $y$ axes. This Bloch function solves the eigenvalue problem $b\phi_{k,n} = [(\boldsymbol{\nabla} + i\mathbf{k})^2/2 + R(r)]\phi_{k,n}$, where $\boldsymbol{\nabla} = (\partial/\partial x, \partial/\partial y)$, $R(r)$ is the profile of the periodic array, and $b$ is the propagation constant. For the SSH array, the topological phase corresponds to $\Delta > 0$ with $P_x = P_y = 1/2$ and the trivial phase is identified by vanishing polarizations $P_x = P_y = 0$ for $\Delta < 0$ [82-84].

We start elucidating the linear properties of the structure, as they are central for understanding of the topological properties of the system. First, we consider a truncated 2D SSH array with 49 unit cells with different dimerizations $\Delta$. The linear eigenmodes propagating in the $z$-direction can be found as $\psi(x, y, z, t) = w(x, y) \exp(ibz)$ where $b$ is the eigenvalue (propagation constant) of the mode, while a real time-independent function $w$ describes the mode profile in the transverse plane. The dependence of the eigenvalues $b$ on the dimerization parameter $\Delta$ is shown in Fig. 1(a) for $d = 3$. One can see that for positive $\Delta$, corner states appear in the topological gap between first and second bands in accordance with nonzero values of bulk polarizations. The corresponding corner state is highlighted in red. The bottom row of the figure shows examples of linear modes in the topological and nontopological phases. The localization of the topological corner modes progressively increases with the increase of $\Delta$ [see Figs. 1(d) and 1(e)]. In addition to corner states, one can see the appearance of the group of eigenvalues associated with edge states, which we will further call the *edge state band* [an exemplary profile of such edge state is presented in Fig. 1(f)]. In the nontopological regime, at $\Delta < 0$ all modes of the array are delocalized; one illustrative example is presented in Fig. 1(g). Below, we will use topological corner states at $\Delta = 0.75$ to construct nonlinear 3D light bullets. The corner states are degenerate for the large array considered here, thus one can always select linear combinations of such states localized in only one corner of the structure, as in Fig. 1(d). Topological protection of linear corner states has been tested by adding small disorder into waveguide depths and positions, which did not lead to appreciable shifts of their propagation constants. Note that the structure considered here is large enough, with a well-developed gap, as seen from comparison with the spectrum of a larger array with 196 cells shown in Appendix A.

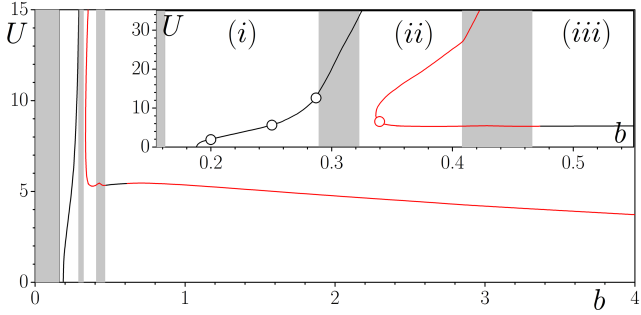

Fig. 2. Family of light bullets bifurcating from linear topological corner modes in a 2D SSH array with $\Delta = 0.75$ and $d = 3$. Black branches are stable, while red ones are unstable. Shaded regions show bulk and edge state bands. In the linear limit, where the propagation constant $b$ tends to a linear propagation constant [red line in Fig. 1(a)] the bullet energy $U$ and the amplitude $a_s$ vanish. The inset illustrates an enlarged image of such a dependence. The dots correspond to the bullets depicted in Fig. 3.

To search for nonlinear bifurcation of the family of light bullets from linear topological corner states we consider stationary soliton profiles in the form $\psi(x, y, z, t) = w(x, y, t) \exp(ibz)$. Substitution of this expression in Eq. (1) leads to the nonlinear problem

$$bw = \frac{1}{2}\left(\frac{\partial^2 w}{\partial x^2} + \frac{\partial^2 w}{\partial y^2}\right) + \frac{1}{2}\frac{\partial^2 w}{\partial t^2} + w^3 + R(x,y)w, \qquad (3)$$

where $b$ is a propagation constant that defines the energy of the bullet $U = \iiint |w|^2 \, dxdydt$, its amplitude and width. We solved this equation using a modified squared operator method [85]. The dependence of $U$ on $b$ that we found is shown in Fig. 2. The gray areas in this figure correspond either to the allowed bands of the linear spectrum or to the band, occupied by the edge states. There are three different regions in the figure: *(i)* the part of the topological gap below the band of edge states, *(ii)* a topological gap above the band of edge states, and *(iii)* a semi-infinite gap. Remarkably, because light bullets bifurcate from localized corner modes, their energy vanishes at the $b$ value corresponding to the propagation constant of the corner state. When $b$ approaches the bifurcation point, the amplitude of the bullet decreases, its spatial width approaches the width of the corner topological mode, while the temporal width drastically increases.

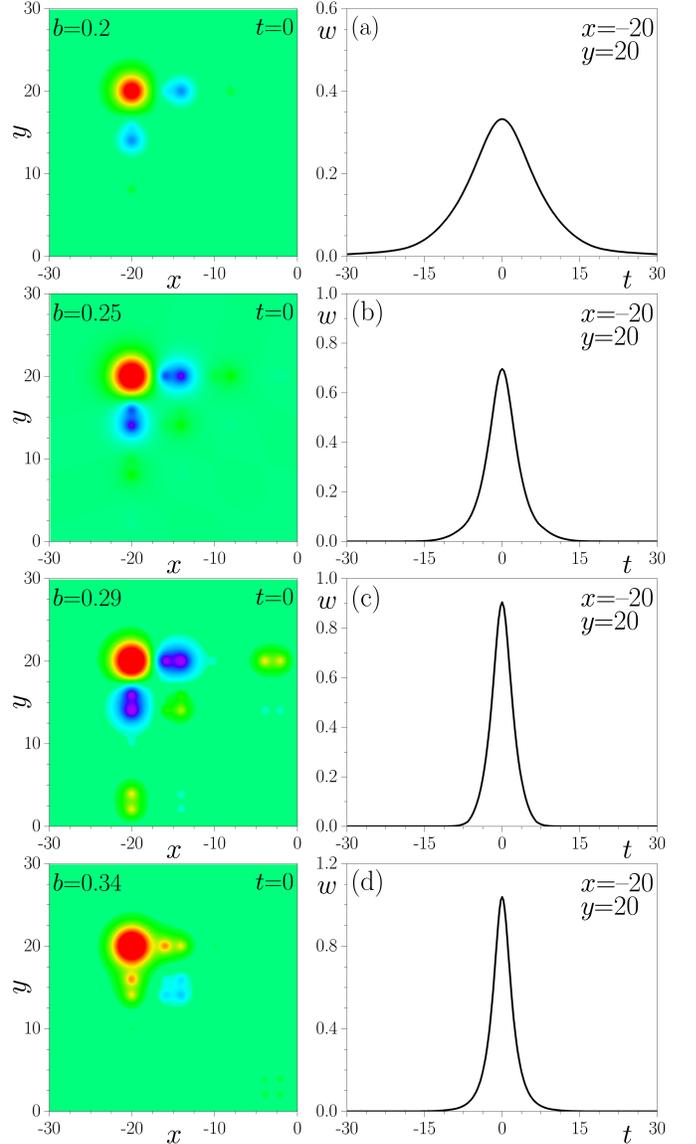

Fig. 3. Spatial distributions at $t = 0$ (left column), and temporal profiles at $x = -20$, $y = 20$ (right column) for light bullets with (a) $b = 0.2$, (b) $b = 0.25$, (c) $b = 0.29$, (d) $b = 0.34$, corresponding to the dots in Fig 2. In the left column, the red regions correspond to high intensities and magenta regions correspond to low intensities. Light bullets from gap *(i)* have the structure of tails associated to topological states, while in gap *(ii)* the corner bullet acquires in-phase tails. The temporal width of the bullets increases when $b$ approaches the bifurcation point from the corner mode. Here $\Delta = 0.75$.

In Fig. 3 we show examples of the cross-sections at $t = 0$ and $x \approx -20$, $y \approx 20$ of solutions from parts *(i)* [Fig. 3(a), (b) and (c)] and *(ii)* [Fig. 3(d)] of the topological gap. Increasing the nonlinearity may drive the corner bullets into the band of edge states, causing their coupling with edge modes. One can see that in this case the bullet acquires long tails along the edges of the array [see Fig. 3(b) and (c)].

Noteworthy, bullets from region *(i)* have an out-of-phase tail in neighboring waveguides to the corner one [see Fig. 3(d)], while the tails of light bullet from regions *(ii)* and *(iii)* are in-phase [see Fig. 3(a)-(c)]. The temporal width increases as the propagation constant decreases, as shown in the right column of the figure. We also found that in region *(ii)*, the family is divided into two branches. In the semi-infinite gap, as the propagation constant $b$ increases, the amplitude of the topological light bullet increases, while the spatial and temporal widths decrease.

We checked the stability of the 3D states via comprehensive propagation simulations. We expect that the solitons corresponding to the stable branches are able to withstand small perturbations without collapse, whereas linearly unstable solitons are expected to either collapse or spread, depending on the type and strength of the perturbation. We simulated the evolution of perturbed solitons using the input conditions $\psi(z=0) = w(x,y,t)(1 + n_{re} + in_{im})$, where $n_{re}$ and $n_{im}$ represent a small noise, whose amplitude is uniformly distributed in the interval $[-0.05, 0.05]$. We found that stability properties in semi-infinite gap agrees with the Vakhitov-Kolokolov criterion [86] implying stability for $\partial U/\partial b > 0$ branches and instability for branches with negative slope. In contrast to our 3D solitons, 2D (1D) nonlinear corner (edge) states in the SSH array exhibit stability throughout the semi-infinite gap. Stable 3D states in Fig. 2 are shown in black, while unstable states are shown in red. Both branches of region *(ii)* are unstable, however, it should be noted that for larger depth $p$ of waveguides, it is possible to obtain stable states from this gap too. Remarkably, the entire branch of bullets bifurcating from the linear corner state corresponds to stable states, even when it penetrates into the edge state band. This suggests that higher-order topological insulators may allow to observe stable bullets even with low energies/peak amplitudes. This is in contrast to usual nontopological waveguide arrays, where stable 3D states exist only in narrow band of energies, limited both from below and above.

Examples of stable and unstable evolution of topological light bullets are shown in Fig. 4. The left column of the figure shows the evolution of the peak amplitude $a_s$ of the bullets along propagation, while the right column shows representative isosurfaces ($|\psi| = $ const) at various $z$. The amplitude, temporal and spatial widths of stable perturbed light bullets belonging to region *(i)* oscillate only slightly, and no collapse or breakup is observed during propagation over considerable distances [black line in Fig. 4(a)]. In contrast, under the action of small noise unstable solitons usually quickly decay, as shown in Fig. 4(b).

In practice, the impact of higher-order effects is critical for the excitation of long-lived light bullets. Such effects occur with ultrashort pulses and quickly destroy the bullets. However, as the light bullets described here are stable at low energies, one may excite them with relatively large temporal width. For instance, a temporal full width at half maximum (FWHM) of the topological bullet with $b = 0.2$ shown in Fig. 3(a) is $14.2$, which corresponds to a time duration of about $115$ fs, and the light bullet with $b = 0.19$ has a FWHM of $22.6$ ($183$ fs). On the other hand, the spatial localization of the bullets is dictated by the localization degree of the topological corner state from which they bifurcate (see Appendix B), and one can readily control such degree by changing the dimerization $\Delta$ of the array or the waveguide depth $p$.

To elucidate the robustness of our light bullets we also examined how disorder in the underlying array impacts their propagation. To such end, we consider the propagation of one of the light bullets from the topological gap in the SSH lattice with diagonal and off-diagonal uncorrelated disorder. We assumed that the depths of waveguides take random values uniformly distributed in the interval $[p(1-\delta_p), p(1+\delta_p)]$. We also changed the spatial position of each waveguide by a random shift uniformly distributed within $[-\delta_d, \delta_d]$ along the $x$ and $y$ axes. Such a disorder broadens the bulk bands and simultaneously leads to small fluctuations of the propagation constant of the corner states. The evolution of the light bullet peak amplitude versus propagation distance for several disorder realizations is depicted by red curves in Fig. 4 (a) for $\delta_p = 0.03$ and $\delta_d = 0.1$. One concludes that stable states are only weakly affected by disorder in the array. Similar conclusions about robustness of the light bullets were obtained by considering $z$-dependent diagonal and off-diagonal disorder that varies with $z$ not too fast, at the scales substantially exceeding $z = 1$.

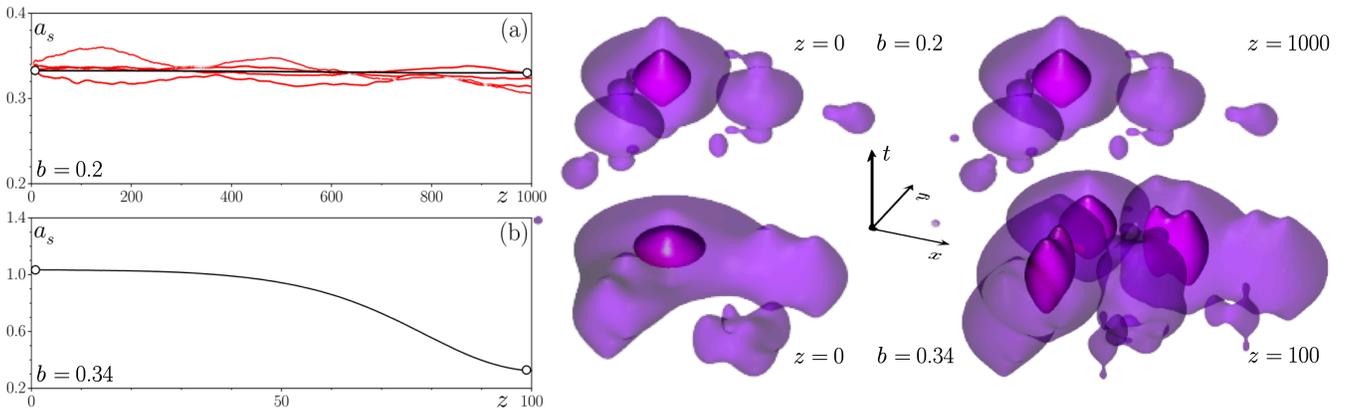

Fig. 4. Evolution of light bullets for $b = 0.2$ (a) and $b = 0.34$ (b). The case $b = 0.34$ represents an unstable state, while $b = 0.2$ is stable. The evolution of the peak amplitude of the light bullets during propagation is shown on the left, and the isosurfaces ($|\psi| = const$) at $z = 100$ and $z = 1000$ for the unstable and stable cases, respectively are shown on the right. The contours are taken at $1/5$ and $1/60$ of the maximum amplitude $a_s$. Red curves in (a) show peak amplitude as a function of propagation distance $z$ for $b = 0.2$ for different realizations of disorder with $\delta_p = 0.03$ and $\delta_d = 0.1$. Here $\Delta = 0.75$.

In summary, we have shown the existence of stable 3D spatiotemporal solitons in a higher-order topological insulator constructed using a 2D Su–Schrieffer–Heeger optical lattice with parameters that are experimentally realizable. We verified the robustness of the 3D states via numerical propagations in the presence of disorder. The result puts forward a different approach to tackle the longstanding problem of experimental formation of long-lived nondiffracting and nondispersing self-sustained 3D objects in nonlinear optical media, by exciting them as corner states in a photonic topological insulator.

**ACKNOWLEDGEMENTS**

This research was funded by the Russian Science Foundation under Grant No. 21-12-00096 and partially by Research Project No. FFUU-2021-0003 of the Institute of Spectroscopy of the Russian Academy of Sciences. Support by Agencia Estatal de Investigación Grants No. CEX2019-000910-S and No. PGC2018-097035-B-I00 funded by MCIN/AEI/10.13039/501100011033/FEDER, Departament de Recerca i Universitats de la Generalitat de Catalunya (Grant No. 47Y48R6YK), and Generalitat de Catalunya (CERCA), Fundacio Cellex and Fundacio Mir-Puig, is acknowledged.

**APPENDIX A: LINEAR SPECTRUM OF 2D SU–SCHRIEFFER–HEEGER WAVE-GUIDE ARRAY**

In the main text of this paper, we use the SSH array with 49 unit cells. To stress that the band-gap structure is similar in larger arrays and that the topological gap is already well-developed in the array with the 49 unit cells considered in the main text, in Fig. 5 we present the spectrum (propagation constants of all linear eigenmodes of array versus dimerization parameter $\Delta$) for a larger array with 196 unit cells. The red dots in this spectrum correspond to corner states appearing between the first and second bulk bands at $\Delta > 0$. They may partially overlap with bulk band at $\Delta < 0.6$ as observed also in [55]. In addition to corner states, one can see the appearance of the group of eigenvalues associated with edge states (blue dots). Black dots correspond to delocalized bulk modes. It can be seen that the boundaries of the gaps and bands from Fig. 1(a) of the main text agree well with the boundaries in the spectrum of larger array from Fig. 5, i.e., the gap is well developed for the array used in the main text.

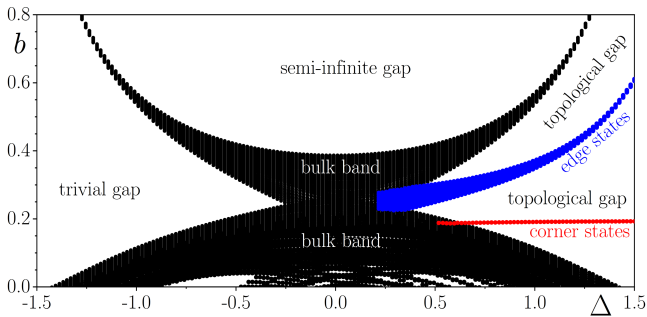

Fig. 5. Propagation constants $b$ of the linear modes of the 2D SSH array vs dimerization parameters $\Delta$. Red dots correspond to corner topological modes, blue dots correspond to edge modes, and black dots correspond to bulk modes.

**APPENDIX B: SPATIAL DISTRIBUTIONS OF LIGHT BULLETS FOR DIFFERENT $\Delta$**

Next, we consider three-dimensional solitons that bifurcate from linear corner states for different dimerization parameters $\Delta$. In the main text, we show that at sufficiently low energies, when $b$ is close to the bifurcation point, the temporal width of the soliton drastically increases. Here we confirm that in this regime the spatial width of the light bullet can be controlled by dimerization parameter $\Delta$. In Fig. 6 we show the examples of the field distributions at $t = 0$ in light bullets with a fixed propagation constant $b = 0.2$ for different $\Delta$ values. One can see that the spatial width of the bullet decreases with increasing dimerization parameter. At the same time, one can clearly see that the field changes its sign in waveguides belonging to different unit cells, which is a characteristic feature for these states of topological origin with propagation constants belonging to the topological gap that clearly distinguishes them from conventional light bullets with in-phase tails.

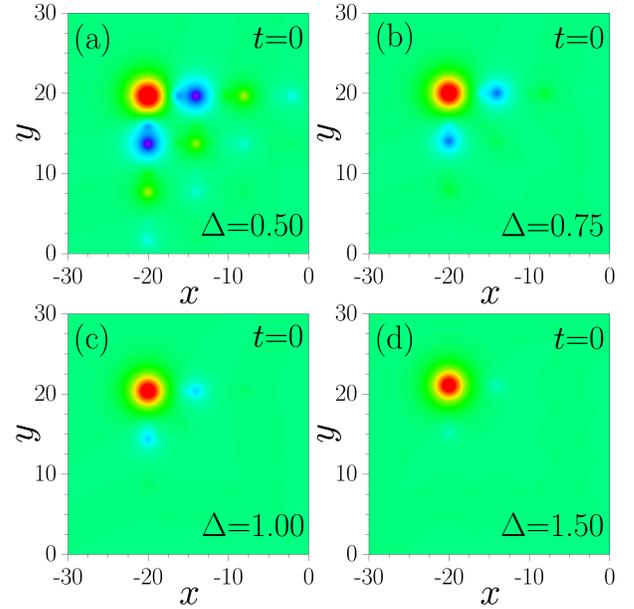

Fig. 6. Spatial field distributions at $t = 0$ in light bullets with (a) $\Delta = 0.5$, (b) $\Delta = 0.75$, (c) $\Delta = 1$, and (d) $\Delta = 1.50$ with fixed propagation constant $b = 0.2$. These light bullets have staggered structure of tails representative for topological states. With increase of $\Delta$ the spatial width of the bullet gradually approaches the width of the linear topological corner mode.